%% file: csd.tex
\documentclass[preprint, letterpaper, nobibnotes, aps, superscriptaddress,endfloats*,prb]{revtex4-1}

\input{preamble}

\begin{document}

\title{Computational science and re-discovery: open-source implementations of ellipsoidal harmonics for problems in potential theory}
\author{Jaydeep P. Bardhan}
\affiliation{Dept. of Molecular Biophysics and Physiology, Rush University Medical Center, Chicago IL 60612}
\author{Matthew G. Knepley}
\affiliation{Computation Institute, The University of Chicago, Chicago IL 60637}

\input{abstract}

\maketitle


\input{brief}

\bibliographystyle{unsrt}
\bibliography{ellipsoidal}

\end{document}

%% file: preamble.tex
\usepackage{graphicx}
\usepackage{amssymb}
\usepackage{amsmath}
\usepackage{listings}
\usepackage{color}
\usepackage{url}
\usepackage{graphicx}        
\usepackage{todonotes}

\def\vr{{\bf r}}

\newcommand{\arxiv}{\let\arxivFig\matt}

\def\XXint#1#2#3{{\setbox0=\hbox{$#1{#2#3}{\int}$}
     \vcenter{\hbox{$#2#3$}}\kern-.5\wd0}}

%% file: abstract.tex
\begin{abstract}
We present two open-source (BSD) implementations of ellipsoidal
harmonic expansions for solving problems of potential theory using
separation of variables.  Ellipsoidal harmonics are used surprisingly
infrequently, considering their substantial value for problems ranging
in scale from molecules to the entire solar system.  In this article,
we suggest two possible reasons for the paucity relative to spherical
harmonics.  The first is essentially historical---ellipsoidal
harmonics developed during the late 19th century and early 20th, when
it was found that only the lowest-order harmonics are expressible in
closed form. Each higher-order term requires the solution of an
eigenvalue problem, and tedious manual computation seems to have
discouraged applications and theoretical studies.  The second
explanation is practical: even with modern computers and accurate
eigenvalue algorithms, expansions in ellipsoidal harmonics are
significantly more challenging to compute than those in Cartesian or
spherical coordinates.  The present implementations reduce the
``barrier to entry'' by providing an easy and free way for the
community to begin using ellipsoidal harmonics in actual research.  We
demonstrate our implementation using the specific and physiologically
crucial problem of how charged proteins interact with their
environment, and ask: what other analytical tools await re-discovery
in an era of inexpensive computation?
\end{abstract}
%
%
%

%% file: brief.tex
\section{Introduction}\label{sec:introduction}

It is intuitively obvious that a cow is much better approximated by an
ellipsoid than by a sphere.  Less obvious, or at least less well
known, is the method of \textit{ellipsoidal harmonics} for solving the
ellipsoidal cow exactly~\cite{Hobson31,DassiosBook}.  In an effort to
increase the popularity and impact of these fascinating functions, we
present in this paper two open-source (BSD) implementations for
calculating the ellipsoidal harmonics and solving problems of
potential theory (they may be downloaded freely
online~\cite{bitbucket-ellipsoidal}).  To demonstrate the methods'
correctness, we model the physiologically crucial electrostatic
interactions between a protein and surrounding water using a popular
continuum theory based on the Poisson equation~\cite{Sharp90,Honig95}.
Our results indicate, that improved numerical methods are needed, and
we hope that releasing these codes will not only encourage application
scientists to try ellipsoidal harmonics, but also motivate numerical
analysts to help develop more accurate computational approaches.

Our implementations, written in Python and MATLAB~\cite{Matlab},
employ a variety of insights developed over more than a century of
research~\cite{Hobson31,Romain01,Xue11}; many important results were
published during the mid-19th century and early 20th by greats like
Heine~\cite{Heine78}, and Charles Darwin's son Sir George Howard
Darwin~\cite{Darwin01}.  From the authors' perspective, these
surprisingly direct ties to mathematical history offer an unusual and
inspiring aspect to their study.  Ellipsoidal coordinates are the most
general coordinate system for which the Laplace equation is separable,
as the other coordinate systems for which Laplace is separable can all
be considered degenerate forms of ellipsoidal
coordinates~\cite{MorseFeshbach53}.  In fact, the functions for each
coordinate direction satisfy the same scalar equation, the Lam\'e
equation.  It seems somewhat ironic that the most general separable
coordinate system should lead to this simple Cartesian-like result
rather than to a more complicated form as found e.g. in spherical
harmonics; the rabbit hole goes much deeper, but for now it suffices
to observe that ellipsoidal harmonics are not a simple generalization
of spherical harmonics.

Unsurprisingly, the more general shape allows ellipsoidal-harmonic
expansions to be more accurate than ones based on spherical
harmonics~\cite{Stiles79,Romain01,Yu03}.  Successful applications
cover most of potential theory: gravity~\cite{Darwin01,Romain01},
electrostatics~\cite{Perram76,Rinaldi82,Rinaldi92,Ford92,Xue11,Kang07},
electromagnetics~\cite{Dassios80,Dassios87,Dassios03,Kariotou04},
hydrodynamics~\cite{Woessner62,Douglas95,Youngren75_2,Ryabov06}, and
elasticity~\cite{Ammari07}.  Specific uses in molecular and biological
sciences include solving the Schr\"{o}dinger
equation~\cite{Levitina00}, modeling van der Waals (close packing)
interactions between molecules~\cite{Stiles79}, design of magnetic
resonance imaging (MRI) devices~\cite{Crozier02} and analysis of
clinical electroencephalography (EEG) and magnetoencephalography (MEG)
data~\cite{Dassios03,Kariotou04}.  Unfortunately, despite the range of
applications and their advantages over spheres, a comparatively small
number of publications actually use, or encourage the use of,
ellipsoidal harmonics in practice.  Notable and welcome exceptions may
be found in Sten's excellent tutorial review~\cite{Sten06} and
presumably in the upcoming book of Dassios~\cite{DassiosBook}, who has
been one of the subject's leading developers and
champions~\cite{Dassios80,Dassios02,Dassios03,Dassios09,Dassios09_2}. A
few older books also address the theory~\cite{Bowman,Byerly}, but for
decades the authoritative reference has been E. W. Hobson's 1931
text~\cite{Hobson31}, published only two years before he passed away
(for a charming obituary, see~\cite{HobsonObituary}).

One reason for the relative dearth of papers using ellipsoidal
harmonics is that they are hard to compute: unfortunately for
investigators who might be interested in actually using ellipsoidal
harmonics, widely available references do not address the numerous
critical challenges associated with actually computing ellipsoidal
harmonics to arbitrary order.  Na\"{i}ve numerical algorithms for
their determination are unstable~\cite{Sona95}, and careful
reformulations are
required~\cite{Rinaldi82,Dobner97,Ritter98,Dobner98,Romain01}.  This
difficulty contrasts sharply to calculations in spherical harmonics or
Cartesian space, where the functions for the expansion are given in
closed form and can be computed easily.  A second possible
explanation, more speculative, is that the foundations of ellipsoidal
harmonic analysis were developed many decades before the advent of
digital computers~\cite{Niven1891,Darwin01,Hobson31}; actually using
ellipsoidal harmonics to arbitrary order necessitates extensive
computation, whether manual or by computer.  Computation is necessary
because the harmonics are defined as polynomials whose coefficients
are solutions to eigenvalue problems.  As such only the lowest-order
modes can be expressed in closed form~\cite{Hobson31}, and all other
modes must be determined numerically (by the Abel-Ruffini theorem).
Furthermore, the harmonics depend on the semi-axes of the ellipsoid,
necessitating a new set of tedious manual computations for every
problem.  For that matter, we note that even the modes available in
closed form require cumbersome algebraic
manipulations~\cite{Dassios03,Dassios09_2}.  It seems that the large
quantities of manual arithmetic led practitioners in many fields to
adopt less accurate, but much more easily computed, spherical harmonic
methods.

Thus, it seems possible that open-source implementations of
ellipsoidal harmonics, based on the latest numerical algorithms for
their computation, could encourage their re-discovery and wide
application in computational science.  The present work offers a
simple demonstration that deriving an analytical solution using
ellipsoidal harmonics is not any harder than deriving one using
spherical harmonics.  Instead, the challenge is in computing the
functions themselves, a problem met by the released implementations;
though, again, we acknowledge that a number of improvements are
needed.  We begin the technical portion of the paper in
Section~\ref{sec:theory}, which presents the application setting
(continuum electrostatics for biomolecular solvation) and the
mathematical components needed to compute ellipsoidal harmonics.  In
Section~\ref{sec:implementation} we describe our implementations at a
high-level but highlight critical details and subtle issues.
Section~\ref{sec:results} presents computational results, and
Section~\ref{sec:discussion} concludes the paper with a discussion of
remaining challenges and possible future directions.

\section{Theory}\label{sec:theory}
Space constraints preclude full accounting of these topics; we present
only the most relevant formulae and refer interested readers to the
extensive and excellent treatments which have made our work
possible~\cite{Hobson31,Romain01,Dassios03,Xue11}.  It is unfortunate
that the clearest publications do not use a single, consistent
notation; some types of analyses and identities are more readily seen
in one notation or another.  We have attempted to use the most popular
conventions where possible, and apologize for introducing yet another
mongrel notation.

\subsection{Biomolecule Electrostatics}\label{subsec:biomolecule-electrostatics}
We assume that the biomolecule is a triaxial ellipsoid whose surface
satisfies
\begin{equation}
  \frac{x^2}{a^2} + \frac{y^2}{b^2} + \frac{z^2}{c^2} = 1,\label{eq:ellipsoid-definition}
  \end{equation}
with $0 < c < b < a < \infty$ called the ellipsoid's semiaxes.  In
this paper, we use the term \textit{ellipsoid} to mean a tri-axial
ellipsoid (three unequal semi-axes), which is the most general
ellipsoidal geometry.  Ellipsoids with two equal semi-axes are called
\textit{spheroids}, these special cases are still more more
straightforward to
study~\cite{Ahner86,Ahner94,Ahner94_2,Ahner96,Langebartel91,Martinec97,Jansen00}
and apply~\cite{AmbiaGarrido08,DiBiasio10,Frankel12}.

The molecular interior is assumed to be a homogeneous local dielectric
of relative permittivity $\epsilon_1$ obeying linear response, and the
molecular charge distribution is (without loss of generality) assumed
to be a set of $Q$ discrete point charges, the $i$th of which being
situated at $\vr_i$ and having value $q_i$. Thus, inside the molecule
(called region 1), the electrostatic potential obeys the Poisson
equation
\begin{equation}
\nabla^2 \Phi_1(\vr) = -\frac{1}{\epsilon_1} \sum_{k=1}^Q q_i \delta(\vr-\vr_i) 
  \end{equation}
where $\delta(\vr)$ is the Dirac delta function.  The solvent outside
the molecule (region 2) is modeled as a homogeneous dielectric with
dielectric constant $\epsilon_2$ and no fixed charges (i.e. we are
modeling a salt-free solution), so the potential in this region obeys
the Laplace equation
\begin{equation}
  \nabla^2 \Phi_2(\vr) = 0.
  \end{equation}
At the dielectric boundary, i.e., the ellipsoid surface defined by
Eq.~\ref{eq:ellipsoid-definition}, the potential is continuous, as is
the normal component of the electric displacement field
$\mathbf{D}(\vr) = \epsilon(\vr) \mathbf{E}(\vr)$, so for a point
$\vr_S$ on the surface,
\begin{align}
  \Phi_1(\vr_S) &= \Phi_2(\vr_S) \label{eq:BCPot})\\
  \epsilon_1 \frac{\partial \Phi_1}{\partial n}(\vr_S) & =
  \epsilon_2 \frac{\partial \Phi_2}{\partial n}(\vr_S)\label{eq:BCNormalD}
  \end{align}
and we assume that the normal direction $\hat{\mathbf{n}}(\vr_S)$
points outward into the solvent.  The potential is assumed to decay
sufficiently rapidly as $||\vr|| \rightarrow \infty$.  We note that
more sophisticated electrostatic models have been
presented,~e.g.~\cite{Xue11}; here we use a simple and popular
one~\cite{Sharp90,Honig95} to demonstrate our implementation.

\subsection{Ellipsoidal Coordinates and Separation of Variables}\label{subsec:ellipsoidal-coordinates}

In the ellipsoidal coordinate system, a point denoted in Cartesian
space $\vr = (x, y, z)$ is written as $(\lambda, \mu, \nu)$; each
ellipsoidal coordinate is a root of the cubic
\begin{equation}
  \frac{x^2}{s^2} + \frac{y^2}{s^2-h^2} + \frac{z^2}{s^2-k^2} = 1
  \end{equation}
with
\begin{align}
h^2 &= a^2 - b^2\\
k^2 &= a^2 - c^2
\end{align}
and we take the positive square roots so that $0 < h < k$.  The
squares of the ellipsoidal coordinates are in the ranges
\begin{align}
  \lambda^2 &\in \left[k^2,+\infty\right.\left.\right)\\
    \mu^2 &\in \left[h^2,k^2\right]\\
    \nu^2 &\in \left[0, h^2\right].
  \end{align}
Points on the surface of the ellipsoid with semi-axes $a$, $b$, and
$c$ satisfy $\lambda = a$.  Many texts enforce that $\lambda$ is
positive, i.e. $\lambda \in \left[k, +\infty\right.\left.\right)$, by
  analogy with the radial coordinate in spherical systems.  We have
  found non-negativity assumption problematic for inverse coordinate
  transforms and do not use it (see discussion in
  Sections~\ref{sec:implementation} and~\ref{sec:discussion}).

Our expressions for coordinate transformations come from Romain and
Jean-Pierre~\cite{Romain01}, who note that for a given point
$\vr=(x,y,z)$, the magnitudes of the ellipsoidal
coordinates~$(\lambda,\mu,\nu)$ can be computed via
\begin{align}
\lambda^2 &= 2 \sqrt{Q}\cos\left(\frac{\theta}{3}\right) - \frac{w_1}{3}\label{eq:cart-to-ell}\\ 
\mu^2 &= 2\sqrt{Q}\cos\left(\frac{\theta}{3}+\frac{4\pi}{3}\right) -\frac{w_1}{3}\nonumber\\
\nu^2 &= 2\sqrt{Q}\cos\left(\frac{\theta}{3}+\frac{2\pi}{3}\right) -\frac{w_1}{3}\nonumber
\end{align}
where
\begin{align}
  Q &= \frac{w_1^2-3 w_2}{9} \\
  R &= \frac{9 w_1 w_2 - 27 w_3 - 2 w_1^3}{54}\\
  \cos\theta &= \frac{R}{\sqrt{Q^3}}\\
  w_1 & = -(x^2 + y^2 + z^2 + h^2 + k^2)\\
  w_2 & = x^2 (h^2+k^2) + y^2k^2 + z^2 h^2 + h^2 k^2\\
  w_3 & = -x^2 h^2 k^2.
  \end{align}
The Cartesian coordinates can be computed from the ellipsoidal ones via
\begin{align}
  x^2 & = \frac{\lambda^2 \mu^2 \nu^2}{h^2 k^2}\label{eq:ell-to-cart}\\
  y^2 & = \frac{(\lambda^2-h^2)(\mu^2-h^2)(h^2-\nu^2)}{h^2 (k^2 - h^2)}\nonumber\\
  z^2 & = \frac{(\lambda^2-k^2)(k^2-\mu^2)(k^2-\nu^2)}{k^2 (k^2 - h^2)}\nonumber
  \end{align}
Romain and Jean-Pierre note that the simple Cartesian-to-ellipsoidal
transformation of Eq.~\ref{eq:cart-to-ell} suffers accuracy problems
that can be important in special cases, and present a more
sophisticated approach for improved accuracy~\cite{Romain01}.  Our
present implementation includes only Eq.~\ref{eq:cart-to-ell}, and the
improved algorithm is under development.  All coordinate
transformations using this simpler method are verified by computing
the inverse transform and checking with a tolerance of $10^{-6}$.  Also,
we note that the above expressions correct a small typographical error
in Eq. 7 of their work.

\subsection{The Lam\'{e} Equation and its Solutions}\label{subsec:Lame}

In ellipsoidal harmonics, the Laplace equation separates such that the
solutions for each coordinate satisfy the same differential equation
(Eq.~\ref{eq:Lame}), which is called \textit{Lam\'{e}'s equation}:
\begin{equation}
(s^2 - h^2)(s^2- k^2)\frac{d^2 E}{d s^2}(s)
  +s(2s^2 - h^2 - k^2)\frac{d E}{d s}(s)
  + (p -q s^2) E(s) = 0
  \label{eq:Lame}
  \end{equation}
where $p$ and $q$ are unknown constants.  It turns out that $q =
n(n+1)$, and $p$ is an eigenvalue of a finite-dimensional matrix (see
below).

The solutions of interest belong to four classes of polynomials, which
are often labeled $\mathcal{K}(s)$, $\mathcal{L}(s)$,
$\mathcal{M}(s)$, and $\mathcal{N}(s)$.  For every nonnegative
polynomial degree $n$, there are a total of $2n+1$ number of solutions
from these classes.  Specifically, defining $r=n/2$ for $n$ even and
$r=(n-1)/2$ for $n$ odd, we have $r+1$ solutions in the class
$\mathcal{K}_n(s)$, $n-r$ solutions in $\mathcal{L}_n(s)$, $n-r$
solutions in $\mathcal{M}_n(s)$, and $r$ in $\mathcal{N}_n(s)$.  By
analogy with spherical harmonics, these $2n+1$ solutions of degree $n$
are labeled $E_n^p(s)$, with $p$ (since it is individual to each
solution) serving double duty as a dummy index from 0 to $2n+1$.
Convention holds that these solutions are assembled class-by-class in
alphabetical order, i.e. for $0 \le p \le r$, $E_n^{p}(s) =
\mathcal{K}_n^p(s)$, then the next $n-r$ are the solutions
$\mathcal{L}_n^p(s)$, and so forth.

For brevity, we describe how to determine only one of the classes of
solutions; Romain and Jean-Pierre include a complete and detailed
description of all four~\cite{Romain01}, and our software implementation
discussed in Section~\ref{sec:implementation} can be consulted for full
details.  A stable algorithm introduced by Dobner and Ritter~\cite{Dobner97,Dobner98,Ritter98}
begins by writing the solution class as
\begin{align}
  \mathcal{K}_n^p(s) &= \psi_n^{\mathcal{K},p}(s) P_n^{\mathcal{K},p}(s)\label{eq:Ritter-Lame-product-decomposition}
  \end{align}
with
\begin{equation}
  \psi_n^{\mathcal{K},p}(s) = s^{n-2r}
  \end{equation}
and 
\begin{equation}
P_n^{\mathcal{K},p}(s) = \sum_{j=0}^{r} b_j \left(1-\frac{s^2}{h^2}\right)^j.\label{eq:P}
  \end{equation}
The $p$th solution $\mathcal{K}_n^p(s)$ is determined by the
coefficients $b_j$ in Eq.~\ref{eq:P}, which represent the $p$th
eigenvector of a tridiagonal matrix of the form
\begin{equation}
  \left[\begin{array}{cccccc}
      \tilde{d}_0 &\tilde{g}_0 &0 &0 &\cdots &0\\
      \tilde{f}_1 &\tilde{d}_1 &\tilde{g}_1 &0 &\cdots &0\\
      0 & \tilde{f}_2 &\tilde{d}_2 &\tilde{g}_2 &\cdots &0\\
      0 & &&&\ddots&\vdots\\
      0 & 0 & \cdots & 0 & \tilde{f}_r & \tilde{d}_r      
      \end{array}\right],
  \end{equation}
where the nonzero elements $\tilde{f}_i$, $\tilde{d}_i$, and
$\tilde{g}_i$ depend on the solution class, the ellipsoid semi-axes,
$n$, and $r$.

The above solutions are known as Lam\'{e} functions of the first kind;
they diverge as $s \rightarrow \infty$, making it impossible to write
solutions to the Laplace equation in exterior regions using these
functions.  A closely related set of solutions do behave appropriately
in this limit, and are known as Lam\'{e} functions of the second kind;
these only involve the radial-like coordinate $\lambda$ and are
written
\begin{equation}
  F_n^p(\lambda) = (2 n + 1) E_n^p(\lambda) I_n^p(\lambda)\label{eq:second-kind-Lame}
  \end{equation}
where
\begin{equation}\label{eq:defI}
  I_n^p(\lambda) = \int^{\infty}_\lambda \frac{ds}{\left[E_n^p(s)\right]^2\sqrt{s^2-k^2}\sqrt{s^2-h^2}}.
\end{equation}

\subsection{Ellipsoidal Harmonics}\label{subsec:ellipsoidal-harmonics}
For a given degree $n$ and order $p$, the interior solid ellipsoidal
harmonic is defined as
\begin{equation}
\mathbb{E}_n^p(\lambda,\mu,\nu) = E_n^p(\lambda)E_n^p(\mu)E_n^p(\nu),
\end{equation}
the exterior solid ellipsoidal harmonic as
\begin{equation}
  \mathbb{F}_n^p(\lambda,\mu,\nu) = (2 n + 1) \mathbb{E}_n^p(\lambda,\mu,\nu) I_n^p(\lambda),
  \end{equation}
and the surface ellipsoidal harmonic as
\begin{equation}
  \mathcal{E}_n^p(\mu,\nu) = E_n^p(\mu)E_n^p(\nu).
  \end{equation}
The surface harmonics satisfy the orthogonality condition
\begin{equation}
\int\int_{\lambda = a} \mathcal{E}_n^p(\mu,\nu) \mathcal{E}_{n'}^{p'}(\mu,\nu) ds = \gamma_n^p \delta_{nn'}\delta_{pp'}
\end{equation}
where $\gamma_n^p$ is a normalization constant; writing the surface integral more explicitly,
\begin{equation}
\gamma_n^p = \int_0^h \int_h^k \left(\mathcal{E}_n^p(\mu,\nu)\right)^2
\frac{(\mu^2-\nu^2)}{\sqrt{(\mu^2-h^2)(k^2-\mu^2)}\sqrt{(h^2-\nu^2)(k^2-\nu^2)}} d\mu\;d\nu.
  \end{equation}
Then the Coulomb potential at $\vr$ due to a unit charge at $\vr'$
(with $||\vr|| > ||\vr'||$) is expanded in ellipsoidal harmonics as
\begin{equation}
  \frac{1}{||\vr-\vr'||} = \sum_{n=0}^{\infty}\sum_{p=1}^{2n+1} \frac{4 \pi}{2n+1}\frac{1}{\gamma_n^p} \mathbb{E}_n^p(\vr')\mathbb{F}_n^p(\vr)\label{eq:Coulomb-potential},
  \end{equation}
or a similar form if $||\vr||< ||\vr'||$ (see, e.g.,~\cite{Xue11}).
The normal derivative at the ellipsoid surface defined by $\lambda =
a$ is computed as~\cite{Dassios03}
\begin{equation}
  \frac{\partial}{\partial n} = \hat{\mathbf{\lambda}} \cdot \nabla = \frac{b c}{\sqrt{a^2-\mu^2}\sqrt{a^2-\nu^2}} \frac{\partial}{\partial \lambda}
  \end{equation}

\subsection{Series Solution for Biomolecule Electrostatics}\label{subsec:series-solution}
In analogy with the Kirkwood series solution for a spherical
particle~\cite{Kirkwood34}, we formulate our model problem in
ellipsoidal coordinates $\vr = (\lambda, \mu, \nu)$. The potential in
region 1, $\Phi_1(\vr)$, can be written
\begin{equation}\label{eq:V1Pot}
  \Phi_1(\vr) = \sum^Q_{k=1} \frac{q_k}{\epsilon_1 \left|\vr - \vr_k\right|} + \psi(\vr),
\end{equation}
where $\psi(\vr)$ is the reaction potential. We expand $\psi$ in
ellipsoidal harmonics of the first kind (interior harmonics) since
these are valid within region 1:
\begin{equation}\label{eq:ReactionPotentialExpansion}
  \psi(\vr) = \sum^\infty_{n=0} \sum^{2n+1}_{p=1} B_n^p\; \mathbb{E}^p_n(\vr).
\end{equation}
Similarly, the potential in region 2 may be expanded in ellipsoidal
harmonics of the second kind (exterior harmonics), which are regular
as $||\vr|| \rightarrow \infty$:
\begin{equation}\label{eq:Phi2Expansion}
  \Phi_2(\vr) = \sum^\infty_{n=0} \sum^{2n+1}_{p=1} C_n^p\; \mathbb{F}^p_n(\vr).
\end{equation}
To determine the constants appearing in these expansions, we apply the
boundary conditions Eqs.~(\ref{eq:BCPot}) and~(\ref{eq:BCNormalD}) by
relating them to the Coulomb portion of $\Phi_1$, using the fact that
all charges are contained inside the ellipsoid ($\lambda_k <
\lambda$).  Using the Coulomb-potential expansion from
Eq.~\ref{eq:Coulomb-potential},
\begin{eqnarray}
  \sum^Q_{k=1} \frac{q_k}{\epsilon_1 \left|\vr - \vr_k\right|} &=& \sum^Q_{k=1} \frac{q_k}{\epsilon_1}
    \sum^\infty_{n=0} \sum^{2n+1}_{p=1} \frac{4\pi}{2n+1} \frac{1}{\gamma^p_n} \mathbb{F}^p_n(\vr) \mathbb{E}^p_n(\vr_k) \\
  &=& \sum^\infty_{n=0} \sum^{2n+1}_{p=1} \frac{G_n^p}{\epsilon_1} \mathbb{F}^p_n(\vr)
\end{eqnarray}
where
\begin{equation}
  G_n^p = \sum^Q_{k=1} q_k \frac{4\pi}{2n+1} \frac{1}{\gamma^p_n} \mathbb{E}^p_n(\vr_k).
\end{equation}
Now the first boundary condition, Eq.~(\ref{eq:BCPot}), gives us,
after equating each $(n,p)$ term in order for the relation to hold for
all angles, 
\begin{eqnarray}
  \frac{G_n^p}{\epsilon_1} + B_n^p \frac{\mathbb{E}^p_n(\vr_S)}{\mathbb{F}^p_n(\vr_S)} &=& C_n^p \\
  \frac{G_n^p}{\epsilon_1} + B_n^p \frac{E^p_n(a)}{F^p_n(a)} &=& C_n^p \label{eq:Cnm1}
\end{eqnarray}
In order to apply Eq.~(\ref{eq:BCNormalD}), we differentiate each
series term by term and equate them,
\begin{equation}
  \frac{G_n^p}{\epsilon_2}  + \frac{\epsilon_1}{\epsilon_2} \frac{\frac{\partial \mathbb{E}^p_n(\vr)}{\partial\lambda}|_a}{\frac{\partial \mathbb{F}^p_n(\vr)}{\partial\lambda}|_a} B_n^p = C_n^p.
\end{equation}
We can eliminate the $C_n^p$ coefficients to give the reaction field
coefficients $B_n^p$ in terms of the known source charge coefficients
$G_n^p$,
\begin{equation}\label{eq:Bnm}
  B_n^p = \frac{\epsilon_1 - \epsilon_2}{\epsilon_1 \epsilon_2} \frac{F^p_n(a)}{E^p_n(a)} \left(1 - \frac{\epsilon_1}{\epsilon_2} \frac{\widetilde E^p_n(a)}{\widetilde F^p_n(a)}\right)^{-1} G_n^p
\end{equation}
where, following~\cite{Xue11}, we introduce the notation $\widetilde
E^p_n(\lambda)$ for logarithmic derivative which respect to the
argument
\begin{equation}
  \widetilde E^p_n(\lambda) = \frac{1}{E^p_n(\lambda)} \frac{\partial E^p_n(\lambda)}{\partial\lambda}.
\end{equation}

\section{Implementation Details}\label{sec:implementation}
We have implemented ellipsoidal harmonics in both MATLAB~\cite{Matlab}
and Python.  These implementations are released as open-source
software under the Simplified BSD License and available online.
Computation involves four distinct tasks, which we address in turn
below.  We would like to echo Xue and Deng~\cite{Xue11} in praising
Romain and Jean-Pierre~\cite{Romain01} for their comprehensive
discussions and detailed derivations.

\subsection{Coordinate Transformations}\label{subsec:coordinate-transformation}
Our open source implementations follow a layered design, where high
levels use the results of lower layers. We begin with the bijection
between Cartesian coordinates, and ellipsoidal coordinates, from
Eq.~\ref{eq:cart-to-ell}. We test our implementation by taking a brick
of coordinates in each octant, convert these coordinates to
ellipsoidal coordinates, and then convert back and compare to the
original Cartesian coordinates. By testing each octant, we assure that
we have correctly handled the sign ambiguity in Eq.~\ref{eq:cart-to-ell} or
Eq.~\ref{eq:ell-to-cart}.

The sign ambiguities associated with squaring the coordinates in these
equations can be resolved by appealing to the Lam\'e functions of
first order, $\mathbb{E}_1^p(\lambda,\mu,\nu)$. These harmonics are
just the dipoles (e.g. $\mathbb{E}_1^0 \propto x$) , and as such have
the same sign as the traditional dipoles. Denoting the sign of a
coordinate, say $x$, with the definition $s_x = \mathrm{sgn}(x)$,
\begin{align}
  s_x &= \mathrm{sgn}\left(E^0_1(\lambda) E^0_1(\mu) E^0_1(\nu)\right) \\
  s_y &= \mathrm{sgn}\left(E^1_1(\lambda) E^1_1(\mu) E^1_1(\nu)\right) \\
  s_z &= \mathrm{sgn}\left(E^2_1(\lambda) E^2_1(\mu) E^2_1(\nu)\right),
\end{align}
however this just moves the sign ambiguity to the Lam\'e functions. We
have, from Romain Table III or by taking square roots of
Eq.~\ref{eq:ell-to-cart},
\begin{align}
  k h x &= \lambda \mu \nu \\
  h_1 h y &= \sqrt{\lambda^2 - h^2_z} \sqrt{\mu^2 - h^2_z} \sqrt{h^2_z - \nu^2} \\
  h_1 k z &= \sqrt{\lambda^2 - h^2_y} \sqrt{h^2_y - \mu^2} \sqrt{h^2_y - \nu^2},
\end{align}
where the sign of the square roots must be chosen consistently. We
choose the sign of each square root as the product of the sign of
argument and a sign determined by the semifocal axis. Thus we have
\begin{align}\label{eq:sign}
  s_x &= s_\lambda s_\mu s_\nu \\
  s_y &= (s_\lambda s_{h}) (s_\mu s_{h}) (s_\nu s_{h}) = s_\lambda s_\mu s_\nu s_{h} \\
  s_z &= (s_\lambda s_{k}) (s_\mu s_{k}) (s_\nu s_{k}) = s_\lambda s_\mu s_\nu s_{k}.
\end{align}
Multiplying the first and second, and first and third equations,
\begin{align}
  s_x s_y &= s_{h} \\
  s_x s_z &= s_{k}.
\end{align}
Now our system has the solution
\begin{align}\label{eq:elipSign}
  s_\lambda &= s_x s_y s_z \\
  s_\mu  &= s_x s_y \\
  s_\nu  &= s_x s_z,
\end{align}
which we can check using Eq.~\ref{eq:sign},
\begin{align}
  s_x &= (s_x s_y s_z) (s_x s_y) (s_x s_z) \\
  s_y &= s_x (s_x s_y) \\
  s_z &= s_x (s_x s_z).
\end{align}
We use the sign assignment
\begin{align}
  s_\mu &= s_{h} \\
  s_\nu &= s_{k}.
\end{align}
when computing the sign of $\psi^p_n$ used in the Lam\'e functions, and found in Romain Table II. This also gives the
simple formula for signs of Cartesian coordinates,
\begin{align}\label{eq:cartSign}
  s_x &= s_\lambda s_\mu s_\nu \\
  s_y &= s_\lambda s_\nu \\
  s_z &= s_\lambda s_\mu.
\end{align}

\lstset{language=python,basicstyle=\ttfamily}
In Python, the \lstinline!EllipsoidalSystem!
class defines an ellipsoidal coordinate system using $(a, b, c)$, and
accomplishes the conversion using \lstinline!ellipsoidalCoords()! and
\lstinline!cartesianCoords()!. The corresponding functions in MATLAB are \lstinline!approxCartToEll()! and
\lstinline!ellToCart()!. The sign of Cartesian coordinates are
calculated using Eq.~\ref{eq:cartSign}, and the sign of ellipsoidal
coordinates comes from Eq.~\ref{eq:elipSign}. There is also a sign
ambiguity in the Lam\'e functions, which use the sign of
$\sqrt{\lambda^2-h^2}$ and $\sqrt{\lambda^2-k^2}$ (our two signs in the derivation above)
in the evaluation of $\psi$. Thus, the
\lstinline!calcLame()!, also \lstinline!calcLame()! in MATLAB, method also takes the sign of $\mu$ and $\nu$,
regardless of the coordinate used for evaluation.

\subsection{Evaluating the First-Kind Lam\'{e} Functions}\label{subsec:eigenvalue-problem}
The evaluation of Lam\'e functions using is straightforward, following
Romain; in the Python implementation, see the \lstinline!calcLame()!
method. The $(n,p)$ is first converted to a solution classs
($\mathcal{K}$, $\mathcal{L}$, $\mathcal{M}$, or $\mathcal{N}$), and a
class-specific index $p'$ into the solutions in that class.  For
instance, for $n=2$ there exist 5 solutions: 2 $\mathcal{K}_2$ and one
of each of the others, so $(2,2)$ is associated with
$\mathcal{L}_2^0$.

The factor $\psi$ can then be calculated trivially; Romain Table II
gives the factors for all four solution classes~\cite{Romain01}.
Next, we assemble the tridiagonal matrix for the given Lam\'e type and
solve the eigenproblem. Since the matrix is well scaled, the problem
is easily solved using the default LAPACK routine. We use the $p'$th
eigenvector to provide the coefficients for the polynomial of
Eq.~\ref{eq:P}.  Following standard practice, we normalize the
eigenvector so that the highest power of the argument $\lambda$ in the
sum has coefficient unity; this is prescribed in
Dassios~\cite{Dassios03} and elsewhere. Derivatives of Lam\'e
functions simply invoke the product rule for
Eq.~\ref{eq:Ritter-Lame-product-decomposition}, and derivatives of the
appropriate $\psi$ and polynomial $P$ are simple to compute.

\subsection{Evaluating Integrals for Second-Kind Lam\'{e} Functions}\label{subsec:exterior-integrals}

Lam\'e functions of the second kind, given by
Eq.~\ref{eq:second-kind-Lame}, require the calculation of $I^p_n$ from
Eq.~\ref{eq:defI}. Although this quantity can in principle be
calculated from elementary elliptic integrals~\cite{Romain01}, the
decomposition is ill-conditioned and thus we choose to evaluate the
integral numerically. Adaptive quadrature rules, such as
\lstinline!quad()! from Matlab, are quite efficient for this
job. Since the argument appears as the upper limit of integration, the
derivative of $I^p_n$ amounts to a single evaluation of the integrand.

\subsection{Computing the Normalization Constants}\label{subsec:normalization-constants}
Apparently, the greatest numerical challenge is presented by
evaluation of the normalization constant $\gamma^p_n$ used for surface
ellipsoidal harmonics and representing the Coulomb potential. Romain
and Jean-Pierre present a way to evaluate these constants using four
basic elliptic integrals. Each integral has an integrable singularity
at the end point, and thus requires something like a Gauss--Kronrod
quadrature in order to evaluate the integral accurately through the
use of transformations. The simple midpoint quadrature currently used
in the Python implementation converges quite slowly.  Both
implementations have been checked against analytic answers provided by
Dassios~\cite{Dassios03} for $n = 0, 1, 2$; these low-order terms are
correct to at least five significant figures.

A further complication arises in computing the expansion for the
fundamental solution of Laplace's equation in
Eq.~\ref{eq:Coulomb-potential}.  For certain geometries (depending on
the positions of the points and the ellipsoid semi-axes), the value of
$I_n^p$ can become very large, and is balanced by a very small
$\gamma_n^p$. Both quantities are computed numerically, and so this
cancellation can quickly render the computation inaccurate.  As a
result, expansions beyond order 11-12 are often computed to
unsatisfactory accuracy (see Fig.~\ref{fig:potential_one_over_r}),
although the precise breakdown depends on the semiaxes and position of
the evaluation point. It seems clear that this computation should be
reformulated in order to calculate the quotient directly.

\section{Results}\label{sec:results}
We have verified the two implementations against one another, and
validated the code using the extensive list of identities presented by
Dassios and Kariotou~\cite{Dassios03}, which were invaluable during
development and testing.  In this section, we present three sets of
results to illustrate that the calculations are indeed correct.  For
simplicity in exposition, and due to the Python numerical integration
challenges discussed above, results are from the MATLAB
implementation.

\begin{figure}[ht!]
\centering
\resizebox{5.0in}{!}{\includegraphics{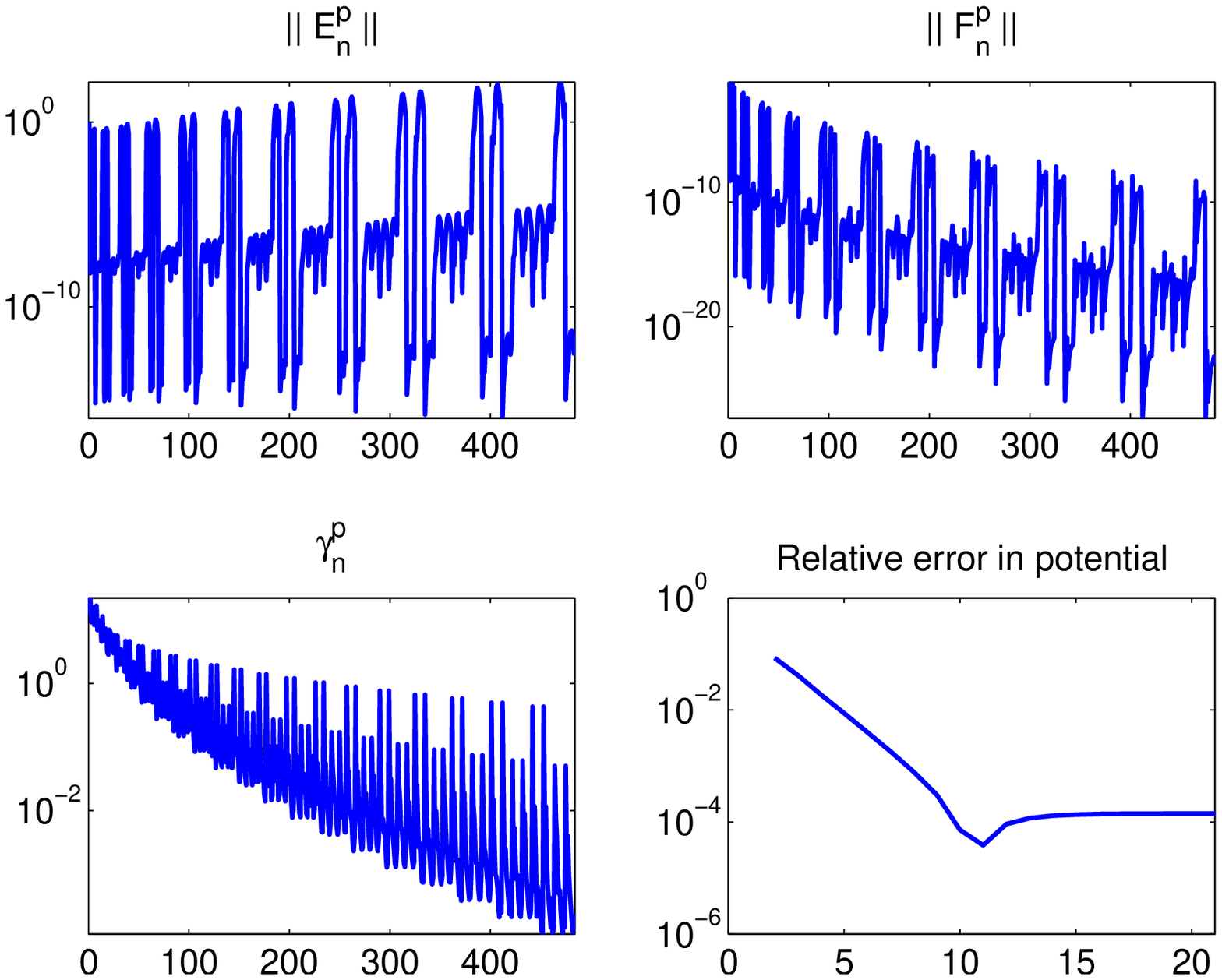}}
\caption{Convergence of the ellipsoidal-harmonic approximation of the
  Coulomb potential as a function of the degree $n$; the ellipsoid is
  defined by $a = 2$, $b = 1.5$, $c = 1$, the source charge is at $(0,
  0, 0.5)$, and the field point is $(0, 0,
  2)$.}\protect\label{fig:potential_one_over_r}
\end{figure}

First, we illustrate that the series expansion of the Coulomb
potential converges to the exact result with increasing order, i.e.
that the error associated with truncating the infinite sum in
Eq.~\ref{eq:Coulomb-potential} goes to zero.
Figure~\ref{fig:potential_one_over_r} plots the magnitude of this
error in the case of ellipsoid with semi-axes $a = 2$, $b = 1.5$, $c =
1$, with internal charge $\vr = (0, 0, 0.5)$ and field point $\vr' =
(0, 0, 2)$; three of the sub-figures illustrate the challenge of
accurate computation, plotting the magnitudes of $\mathbb{E}_n^p$,
$\mathbb{F}_n^p$, and $\gamma_n^p$.  Numerical error in the required
integrations entails a limited accuracy---in this case, a relative
error of approximately $10^{-4}$.  Other test examples exhibit slower
convergence and more complex non-monotonicity.

\begin{figure}[ht!]
\centering
\resizebox{5.0in}{!}{\includegraphics{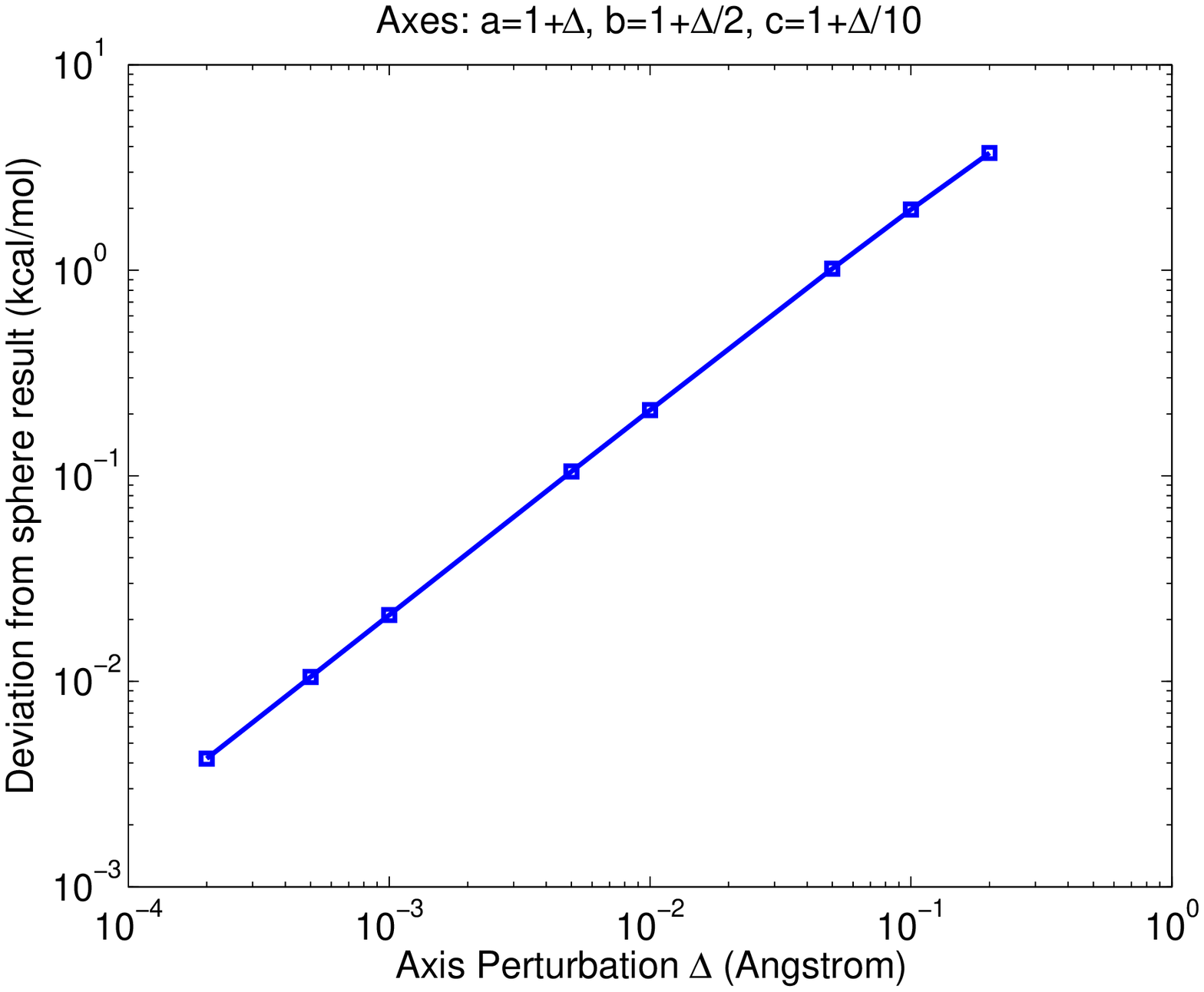}}
\caption{Validating the semi-analytical calculation of electrostatic
  solvation free energies using the analytical result for a sphere
  with a central charge (a quantity known as the Born energy).  We
  define $\Delta > 0$ and an ellipsoid with semi-axes $a = 1 +
  \Delta$, $b = 1 + \Delta/5$, and $c = 1 + \Delta/10$.  A single
  $+1e$ charge is situated at the origin, and we have $\epsilon_1 = 4$
  and $\epsilon_2 = 80$.  As $\Delta \rightarrow 0$, the
  semi-analytical ellipsoidal results converge to the exact energy for
  the sphere, which supports the correctness of our
  implementation.}\protect\label{fig:sphere-convergence}
\end{figure}

Our second result demonstrates that the electrostatic solvation free
energy of a single charge at the origin recovers the well-known result
for the sphere limit, also known as the Born energy, as we model an
ellipsoid with semi-axes $a=1+\Delta$, $b=1+\Delta/5$, $c=1+\Delta/10$
(all units in Angstrom), and let $\Delta$ approach zero
(Figure~\ref{fig:sphere-convergence}).  These calculations show that
the implementations behave correctly for small but finite differences
between the ellipsoid semi-axes. Unfortunately, vanishingly small
differences pose numerical challenges (note that the ranges for $\mu$
and $\nu$ depend on the differences in the semi-axes).  Obtaining the
spherical harmonics in the actual limit requires careful
analysis~\cite{Hobson31,Dassios03}.

\begin{figure}[ht!]
\centering
\resizebox{5.0in}{!}{\includegraphics{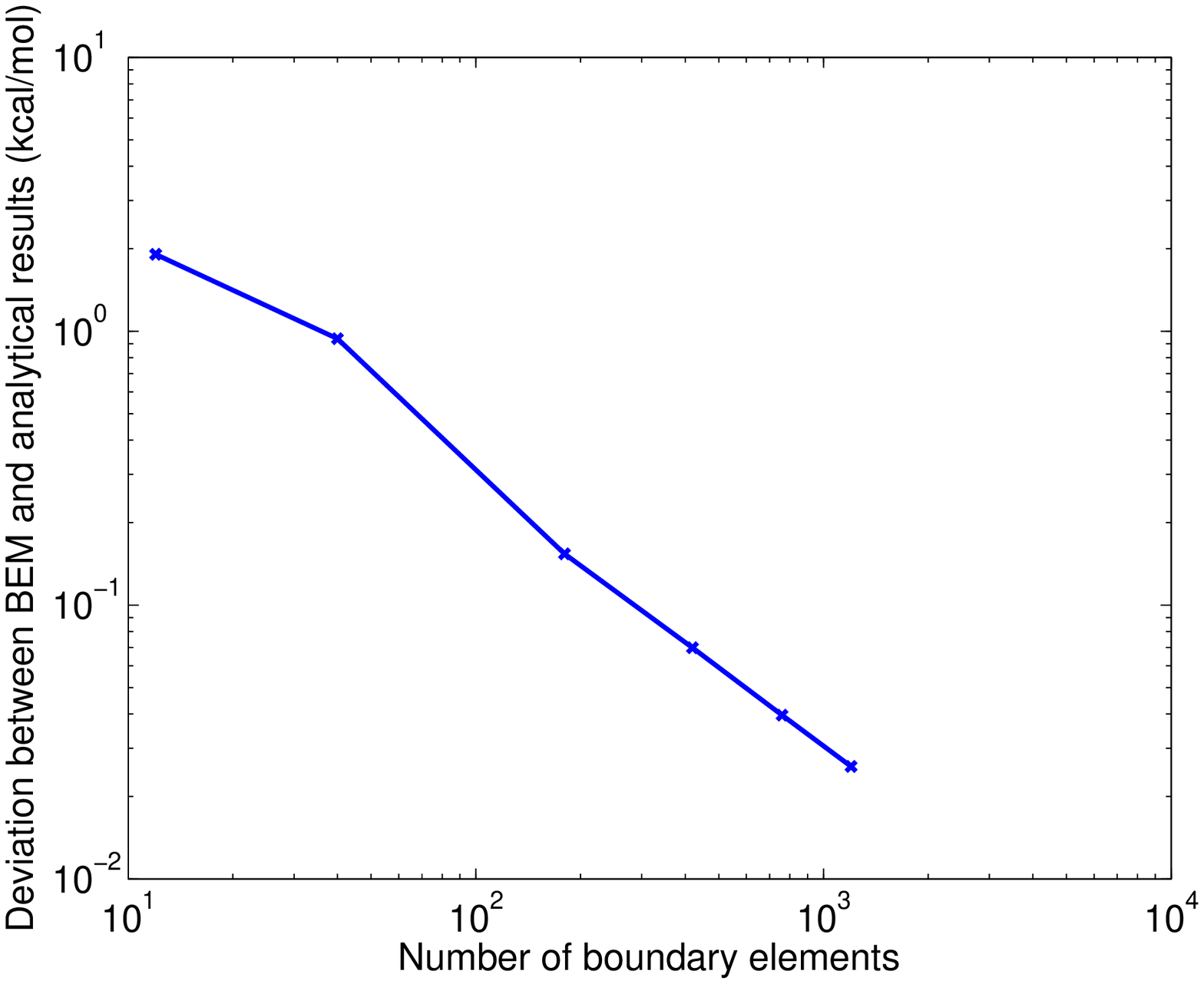}}
\caption{Validating the semi-analytical calculation of electrostatic
  solvation free energies using numerical simulations based on the
  boundary-element method (BEM).  The ellipsoid has semi-axes $a =
  15$, $b = 12$, and $c = 10$, and a single $+1e$ charge is situated
  inside at $(3, 4, 5)$; as in Fig.~\ref{fig:sphere-convergence},
  $\epsilon_1 = 4$ and $\epsilon_2 = 80$.  The BEM results converge
  linearly as a function of the number of unknowns (one per panel),
  which indicates that the semi-analytical method is returning the
  correct result.}\protect\label{fig:bem-convergence}
\end{figure}

Finally, using the same dielectric constants as in the previous
example, we calculate the electrostatic solvation free energy of a
single charge located at $(3, 4, 5)$ inside a protein-like ellipsoid
with semi-axes $a = 15$, $b = 12$, and $c = 10$.  For this system, we
can validate the implementation numerically using simple BEM
calculations, which is also written in MATLAB and included with the
ellipsoidal harmonics software.  Figure~\ref{fig:bem-convergence}
plots the deviation between the semi-analytical result and the
numerical BEM results, as a function of the number of boundary
elements (flat triangles) used to approximate the ellipsoid surface.
We observe the expected linear convergence to the semi-analytical
result, which suggests the correctness of our ellipsoidal-harmonic
calculation.  In the present implementation, more accurate validation
is impractical because meshes are generated from within MATLAB using
the built-in function \texttt{ellipsoid}, and BEM calculations are
performed using dense $O(N^2)$ algorithms.  Fast BEM
solvers~\cite{Lu06,Altman09} will be needed for more stringent
testing.  Also, as Fig.~\ref{fig:potential_one_over_r} illustrates,
more accurate methods need to be found to compute the needed
integrals.

\section{Discussion}\label{sec:discussion}
In this paper we have presented two open-source implementations of
ellipsoidal harmonics to allow rapid, analytical calculations of many
problems of potential theory, including electrostatics,
electromagnetics, elasticity, and fluid mechanics.  The
implementations are written in MATLAB and Python, and released under
the Simplified BSD License, and are freely available online~\ref{bitbucket-ellipsoidal}. These development efforts stemmed from
our interest in improving continuum Poisson-based models of the
electrostatic interactions between biological molecules and the
surrounding solvent~\cite{Sharp90_2,Honig95,Simonson01}.  Many important studies have
relied on analytically solvable spherical
geometries~\cite{Kirkwood34}, and a variety of state-of-the-art
approximate models rely on analyses in spherical harmonics,
e.g.~\cite{Lee02,Sigalov05}.  We hope that the present implementation
of ellipsoidal harmonics, or at least our description of pitfalls we
encountered along the way, may provide a starting point for more
accurate Generalized-Born models~\cite{Lee02,Sigalov05} or other fast
approximations to the Poisson
problem~\cite{Bardhan08_BIBEE,Bardhan09_bounds,Bardhan11_bibee}.

We do not claim to have covered the entire field of ellipsoidal
harmonics, however, or that our implementations are production-level
codes with all mathematical subtleties fully addressed.  The
implementations released with this paper remain under active
development, and interested readers are encouraged to contact us with
questions, requests, bug reports, and suggestions of all kinds.
Currently, the most important area for improvement appears to be in
handling coordinate transformations, particularly methods for
accurately mapping Cartesian to ellipsoidal
coordinates~\cite{Romain01}, and transformations that seem to need
negative $\lambda$ where other publications indicate that non-negative
$\lambda$ is adequate.  Our results also illustrate the need for more
accurate ways to evaluate integrals for the second-kind Lam\'{e}
functions and the normalization constants.

The present implementation includes the basic algorithms needed to
compute the ellipsoidal harmonics, and we have used these primitives
to address a fairly simple problem of potential theory (a
mixed-dielectric Poisson problem).  The primitives can be used for a
variety of other problems in potential theory, which are of interest
on scales ranging from atoms~\cite{Perram76,Rinaldi82} to entire
planets and the solar system~\cite{Darwin01,Romain01}.  In addition,
recent work by Ritter and colleagues has elucidated the relationship
between ellipsoidal harmonics and boundary-integral operators of
potential theory for ellipsoidal boundaries (e.g.,
\cite{Ritter95_spectrum,Ritter98}).  Our approach to approximating
biomolecule electrostatic problems using boundary-integral
methods~\cite{Bardhan08_BIBEE,Bardhan09_bounds,Bardhan11_bibee} should
benefit significantly.  The most recent analysis employed spherical
harmonics and found that accurate approximations could be obtained
using parameters that correspond not to spheres but surfaces close to
spheres~\cite{Bardhan11_bibee}.  We conducted this work because
ellipsoids have been used for a number of other molecular
electrostatic models~\cite{Perram76,Rinaldi82,Rinaldi92,Sigalov06},
and thus the boundary-integral analysis of Ritter et al. may allow new
approach to approximating the Poisson problem.  Also, much of the
earlier work employed only the lowest-order modes (which can be
expressed in closed form), and did not allow extension to the
higher-order ones that must be computed.

In closing: ellipsoidal harmonics have a wide range of applicability,
but historically the mechanical details required to use them have been
an unfortunate deterrent.  Modern computers and numerical algorithms
can easily handle these details to enable powerful new modeling tools, and in
this work we take a first step towards this goal.  It is already clear
that computing these functions can motivate more fundamental numerical
research, and we hope others will find this subject and its
applications equally fascinating.  It seems entirely likely that
mechanical computation derailed the development of other powerful
formalisms in the era before digital computers; perhaps computational
science should undertake a journey of re-discovery.

\section*{Acknowledgements}
The authors thank R. S. Eisenberg for stimulating discussions and for
his support of interdisciplinary collaborations in biology and
computational mathematics.  The work of MGK was supported in part by
the U. S. Army Research Laboratory and the U. S. Army Research Office
under contract/grant number W911NF-09-0488.